\documentclass[twocolumn,times,trackchanges]{aastex63}
\usepackage{mathrsfs}

\shorttitle{Accretion disk reverberation in Mrk~142}
\shortauthors{Cackett et al.}

\begin{document}

\title{Supermassive black holes with high accretion rates in active galactic nuclei. \\
XI. Accretion disk reverberation mapping of Mrk 142}

\correspondingauthor{Edward M. Cackett}
\email{ecackett@wayne.edu}

\author[0000-0002-8294-9281]{Edward M. Cackett}
\affiliation{Wayne State University, Department of Physics \& Astronomy, 666 W Hancock St, Detroit, MI 48201, USA}

\author[0000-0001-9092-8619]{Jonathan Gelbord}
\affiliation{Spectral Sciences Inc., 4 Fourth Avenue, Burlington, MA 01803, USA}

\author[0000-0001-5841-9179]{Yan-Rong Li}
\affiliation{Key Laboratory for Particle Astrophysics, Institute of High Energy Physics, Chinese Academy of Sciences, 19B Yuquan Road, Beijing 100049, People's Republic of China}

\author[0000-0003-1728-0304]{Keith Horne}
\affiliation{SUPA Physics and Astronomy, University of St. Andrews, North Haugh, KY16 9SS, UK}

\author[0000-0001-9449-9268]{Jian-Min Wang}
\affiliation{Key Laboratory for Particle Astrophysics, Institute of High Energy Physics, Chinese Academy of Sciences, 19B Yuquan Road, Beijing 100049, People's Republic of China}
\affiliation{School of Astronomy and Space Sciences, University of Chinese Academy of Sciences, 19A Yuquan Road, Beijing 100049, China}
\affiliation{National Astronomical Observatories of China, Chinese Academy of Sciences, 20A Datun Road, Beijing 100020, China}

\author[0000-0002-3026-0562]{Aaron J. Barth}
\affiliation{Department of Physics and Astronomy, 4129 Frederick
  Reines Hall, University of California, Irvine, CA, 92697-4575, USA}

\author{Jin-Ming Bai}
\affiliation{Yunnan Observatories, Chinese Academy of Sciences, Kunming 650011, People's Republic of China}

\author{Wei-Hao Bian}
\affiliation{Physics Department, Nanjing Normal University, Nanjing 210097, People's Republic of China}

\author{Russell W. Carroll}
\affiliation{Wayne State University, Department of Physics \& Astronomy, 666 W Hancock St, Detroit, MI 48201, USA}

\author{Pu Du}
\affiliation{Key Laboratory for Particle Astrophysics, Institute of High Energy Physics, Chinese Academy of Sciences, 19B Yuquan Road, Beijing 100049, People's Republic of China}

\author[0000-0001-8598-1482]{Rick Edelson}
\affiliation{Department of Astronomy, University of Maryland, College Park, MD 20742-2421, USA}

\author{Michael R. Goad}
\affiliation{Department of Physics and Astronomy, University of Leicester, Leicester, LE1 7RH, UK}

\author[0000-0001-6947-5846]{Luis C. Ho}
\affiliation{Kavli Institute for Astronomy and Astrophysics, Peking University, Beijing 100871, People's Republic of China}
\affiliation{Department of Astronomy, School of Physics, Peking University, Beijing 100871, People's Republic of China}

\author{Chen Hu}
\affiliation{Key Laboratory for Particle Astrophysics, Institute of High Energy Physics, Chinese Academy of Sciences, 19B Yuquan Road, Beijing 100049, People's Republic of China}

\author{Viraja C. Khatu}
\affiliation{Department of Physics \& Astronomy, Western University, London, ON, N6A 3K7, Canada}

\author{Bin Luo}
\affiliation{School of Astronomy and Space Science, Nanjing University, Nanjing, Jiangsu 210093, People's Republic of China}
\affiliation{Key Laboratory of Modern Astronomy and Astrophysics (Nanjing University), Ministry of Education, Nanjing 210093, People's Republic of China}
\affiliation{Collaborative Innovation Center of Modern Astronomy and Space Exploration, Nanjing 210093, People's Republic of China}

\author{Jake Miller}
\affiliation{Wayne State University, Department of Physics \& Astronomy, 666 W Hancock St, Detroit, MI 48201, USA}

\author[0000-0002-7330-4756]{Ye-Fei Yuan}
\affiliation{CAS Key Laboratory for Research in Galaxies and Cosmology, Department of Astronomy, University of Science and Technology of China, Hefei 230026, People's Republic of China}

\begin{abstract}
We performed an intensive accretion disk reverberation mapping campaign on the high accretion rate active galactic nucleus Mrk 142 in early 2019.  Mrk 142 was monitored with the {\it Neil Gehrels Swift Observatory} for 4 months in X-rays and 6 UV/optical filters.  Ground-based photometric monitoring was obtained from the Las Cumbres Observatory, Liverpool Telescope and Dan Zowada Memorial Observatory in {\it ugriz} filters and the Yunnan Astronomical Observatory in {\it V}.  Mrk 142 was highly variable throughout, displaying correlated variability across all wavelengths.  We measure significant time lags between the different wavelength light curves, finding that through the UV and optical the wavelength-dependent lags, $\tau(\lambda)$, generally follow the relation $\tau(\lambda) \propto \lambda^{4/3}$, as expected for the $T\propto R^{-3/4}$ profile of a steady-state optically-thick, geometrically-thin accretion disk, though can also be fit by $\tau(\lambda) \propto \lambda^{2}$, as expected for a slim disk.   The exceptions are the {\it u} and {\it U} band, where an excess lag is observed, as has been observed in other AGN and attributed to continuum emission arising in the broad-line region.  Furthermore, we perform a flux-flux analysis to separate the constant and variable components of the spectral energy distribution, finding that the flux-dependence of the variable component is consistent with the $f_\nu\propto\nu^{1/3}$ spectrum expected for a geometrically-thin accretion disk.  Moreover, the X-ray to UV lag is significantly offset from an extrapolation of the UV/optical trend, with the X-rays showing a poorer correlation with the UV than the UV does with the optical.  The magnitude of the UV/optical lags is consistent with a highly super-Eddington accretion rate.    

\end{abstract}

\section{Introduction} \label{sec:intro}

At typical mass accretion rates onto supermassive black holes in Seyfert galaxies (a few percent of the Eddington limit), accretion is expected to take place via a geometrically thin ($H/R \ll 1$), optically thick accretion disk \citep{shakurasunyaev}.  However, once mass accretion rates exceed the Eddington limit, then radiation pressure becomes important, and is expected to change the structure of the accretion flow. In the `slim disk' class of models, at super-Eddington rates, radiation pressure dominates the accretion flow at most radii, and the disk becomes slim (rather than thin), with $H \lesssim R$ \citep[e.g.][]{abramowicz88}.   Slim disks are characterized by sub-Keplerian rotation and transonic radial motion.  The fast radial transportation in slim disks means that most photons are trapped by optically thick Thomson scattering and advected into the black hole before escaping.  Within this inner photon-trapping region the disk increases significantly in scale height, which can cast a shadow on the outer disk \citep[e.g.][]{wang14_shadow}.  Alternatively, \citet{begelman02} proposes that through the photon bubble instability, the disks may remain thin even above the Eddington limit.  However, observational tests of the nature of super-Eddington accretion flows in Seyfert galaxies are rare.

One way to observationally test the accretion flow and nearby broad-line region (BLR) is to use reverberation mapping \citep[RM;][]{blandmckee82,peterson14}. In RM, time lags between light curves at different wavelengths (either between the continuum and emission lines or the continuum at different wavelengths) can be used to determine the size-scale of the emitting region.  Applying RM to Active Galactic Nuclei (AGN) thought to be accreting at high rates is therefore a way to observationally test super-Eddington accretion.  One class of AGN thought to be accreting at high rates are Narrow-Line Seyfert 1 (NLS1) galaxies. NLS1s are characterized by relatively narrow broad emission lines, strong \ion{Fe}{2} lines, weak [\ion{O}{3}] lines and steep 2--10 keV spectra \citep[e.g.,][]{boller96,veroncetty01}. 

Over the last 7 years or so the Super-Eddington Accreting Massive Black Holes (SEAMBH) collaboration has been performing extensive optical monitoring of super-Eddington AGN candidates that show these characteristics of strong optical Fe {\sc ii} and weak [O {\sc iii}] emission lines \citep[e.g.,][]{Du14,Du16,du18,Hu15}.  These observations show that the BLR structure in these super-Eddington objects differs significantly from more typical sub-Eddington Seyferts \citep{Du16,du18}.  One of the main findings is that these super-Eddington AGN lie below the well known relation between the radius of the H$\beta$-emitting region and the optical luminosity \citep[the $R-L$ relation, e.g.,][]{kaspi00, bentz13}. Hence, this suggests that the BLR size depends on more than just luminosity, i.e. for objects of the same luminosity, those with lower mass and thus higher Eddington ratio show more compact BLRs.  This can be understood as the inner part of the slim disk acting as an optically thick torus, creating a self-shadowing effect that lowers the ionizing flux seen by the BLR \citep{wang14_shadow}.

In order to test the accretion disk structure in a super-Eddington AGN,  we carried out the first accretion disk RM campaign on a super-Eddington AGN, Mrk~142 (PG 1022+519, $z = 0.045$).   Accretion disk RM uses time lags between the continuum at different wavelengths to probe the size and temperature of the accretion disk \citep[e.g.][]{cackett07}.  In the lamp-post reprocessing picture, high energy X-ray/EUV photons from a central corona irradiate the accretion disk, driving variability at longer wavelengths. The hotter, inner disk will respond to variability in the irradiating photons before the cooler, outer disk. This then leads to correlated continuum lightcurves with longer wavelengths lagging shorter wavelengths. Measuring the wavelength-dependence of the lag therefore gives both the size-scale of the disk and its temperature profile for an assumed disk geometry.  For instance, for an optically thick, geometrically thin accretion disk \citep{shakurasunyaev} the temperature profile goes like $T(R) \propto R^{-3/4}$.  Since $\tau \sim R/c$ and $\lambda \propto 1/T$ (from Wien's law), such a temperature profile leads to wavelength-dependent lags following $\tau(\lambda) \propto \lambda^{4/3}$.  On the other hand, since a slim disk has a temperature profile following $T(R) \propto R^{-1/2}$ within the photon-trapping region \citep{wang99}, the wavelength-dependent lags should follow  $\tau(\lambda) \propto \lambda^{2}$ instead.

Recently, advances in accretion disk RM have come from intensive (better than daily) monitoring with the {\it Neil Gehrels Swift Observatory} (hereafter, {\it Swift}) on four Seyferts: NGC~5548 \citep{edelson15,fausnaugh16}, NGC~4151 \citep{edelson17}, NGC~4593 \citep{cackett18,mchardy18} and Mrk~509 \citep{edelson19}. See \citet{edelson19} for a comparison of all four {\it Swift} datasets.  These campaigns have shown three main results.  Firstly, the time lag, $\tau$ generally follows $\lambda^{4/3}$, as expected for a standard thin disk, however, the magnitude of the lags is larger than expected by a factor of 2 -- 3.  Secondly, the lag in the $u$ band (3465\AA) consistently lies above this $\tau \propto \lambda^{4/3}$ relation, which indicates significant continuum emission from the BLR \citep{koristagoad01,koristagoad19,lawther18}.  This is further highlighted by {\it Hubble Space Telescope} monitoring of NGC~4593 which spectroscopically resolved the ``lag spectrum'' in finding a discontinuity at the Balmer jump, as expected if BLR continuum emission is important \citep{cackett18}.  Finally, the X-ray to UV correlation is significantly weaker than the UV to optical correlation \citep{edelson19}, which raises the question of whether the X-rays drive the variability at longer wavelength.  In the case of NGC~5548 the shape of the X-ray lightcurve is not consistent with driving the UV/optical variability \citep{starkey17,gardnerdone17}, while in NGC~4593 it is \citep{mchardy18}.

These four Seyferts all accrete at rates significantly lower than the Eddington limit, and so serve as a good comparison for wavelength-dependent lags measured in objects accreting at much higher rates.  In this paper, we present an intensive accretion disk RM campaign on Mrk~142, using {\it Swift} along with ground-based monitoring.  Mrk~142 has a black hole mass of  $\log(M/M_\odot) = 6.23^{+0.26}_{-0.45}$, and a dimensionless accretion rate of $\dot{\mathscr{M}} = \dot{m}c^2/L_{\rm Edd} = 250$ \citep{li18}.   The paper is organized as follows.  In Section~\ref{sec:obs} we describe the observations, and in Section~\ref{sec:data} we detail the data reduction.  The time series analysis and results are presented in Section~\ref{sec:timeseries}, while in Section~\ref{sec:spec} we use variability to isolate the spectral energy distribution of the disk.  Finally, in Section~\ref{sec:discuss} we discuss the implications.

\section{Observations} \label{sec:obs}

A large, coordinated monitoring campaign on Mrk~142 took place from December 2018 -- June 2019.  The core of the campaign was centered around X-ray and UV/optical observations taken with {\it Swift}.  In addition, we obtained further X-ray observations with {\it NICER}, and supporting ground-based photometric and spectroscopic monitoring from multiple telescope sites.  Details and results from the spectroscopic monitoring and {\it NICER} X-ray analysis will be presented in future follow-up papers.  Here, we focus only on the {\it Swift} and ground-based photometric data.  The ground-based monitoring involved Las Cumbres Observatory (LCO), Liverpool Telescope, Dan Zowada Memorial Observatory (hereafter Zowada Observatory) and the Yunnan Astronomical Observatory.  Further details about observations from each telescope are given below.  A summary of the observations used is given in Tab.~\ref{tab:summary}.

\subsection{Swift}

Mrk~142 was monitored by {\it Swift} from 1 January 2019 to 30 April 2019 through Cycle 14 proposal 1417139 (PI: E.~M.~Cackett).  Initially, {\it Swift} observations were obtained twice per day.  However, following a successful request for Director's Discretionary Time to extend the campaign by 1~month, the cadence of observations became once per day from 20 March 2019 onwards.  In total, 184 epochs of observations were obtained. The typical visit duration was 1000\,s, with the exact length varying depending on scheduling.  X-ray observations were taken in Photon Counting mode.  UVOT exposures were taken in 0X30ED mode, which gives the bluer filters longer exposure times.  For a typical 1000s visit, this gives exposures of approximately 333\,s for {\it UVW2}, 250\,s for {\it UVM2}, 167\,s for {\it UVW1}, and 83\,s for {\it U}, {\it B} and {\it V}.  

\subsection{Las Cumbres Observatory}

LCO is a global network of robotic telescopes.  As part of an LCO Key Project  (KEY-2018B-001, PI: R.~Edelson) monitoring was obtained in Sloan {\it u, g, r, i}, and PanSTARRS {\it z} filters from both the 2-m Faulkes Telescope North at the Haleakala Observatory (OGG), and the 1-m telescope at the McDonald Observatory (ELP). Since most of the ground-based data comes from the two LCO telescopes we adopt the effective wavelengths of the LCO filters\footnote{\url{https://lco.global/observatory/instruments/filters/}} (see Tab.~\ref{tab:lags}) for the subsequent analysis.  On OGG we use the Spectral camera which has a $10.5\arcmin \times 10.5$\arcmin\ field of view, while at ELP we use the Sinistro camera with a $26.5\arcmin\times26.5$\arcmin\ field of view.

Exposures were taken in pairs with individual exposure times being initially 300\,s for {\it u}, 60\,s for {\it g, r} and {\it i}, and 120\,s for {\it z} for OGG.  After analysis of early data the exposure time in the z filter was increased to 240\,s.   For ELP, the initial exposure times were 300\,s for {\it u}, 60\,s for {\it g, r}, and {\it i} and 120\,s for {\it z}.  These exposure times were increased to 600\,s for $u$, 180\,s for $g$, $r$, and $i$ and 360\,s for $z$ after inspection of early data.      Observations with LCO took place between 15 December 2018 and 19 June 2019.

\subsection{Liverpool Telescope}

Photometric monitoring was obtained with the robotic 2-m Liverpool Telescope located on La Palma, Spain through program PL19A01 (PI: M.~Goad).  Observations were taken using the IO:O instrument in $u$, $g$, $r$, $i$, and $z$ filters.  IO:O has $4096 \times 4112$ pixels with a pixel scale of 0\farcs15 per pixel. Pairs of exposures were taken during each epoch, with individual exposure times of 90\,s for $u$, 10\,s for $g$ and $r$, 15\,s for $i$ and 20\,s for $z$.  Observations took place between 3 January 2019 and 22 April 2019.

\subsection{Zowada Observatory}

The Zowada Observatory is a robotic 20-inch f/6.8 PlaneWave telescope located near Rodeo, New Mexico and owned and operated by Wayne State University.  During the monitoring campaign two different detectors were used.  Prior to 19 January 2019 a FLI Proline 16803 CCD with 4096 $\times$ 4096 pixels was used.  On 19 January 2019 a back-illuminated FLI Proline 230-42-1-MB CCD with 2048 $\times$ 2048 pixels was installed.  The pixel size for this detector is 15 microns, leading to a plate scale of 0\farcs9 per pixel.  

Observations began on 31 October 2018 and continued daily (when possible) until 30 May 2019.  Images were obtained using {\it u, g, r, i}, and $z$ filters.    Individual exposure times were 300\,s for $u$, 200\,s for $z$ and 100\,s for {\it g, r}, and $i$.  Multiple exposures per filter were obtained on each night (typically 5 per filter, but it varied depending on weather).  

\subsection{Yunnan Astronomical Observatory}

Observations at the Lijiang Station of the Yunnan Observatories, Chinese Academy of Sciences, were obtained with the 2.4-m telescope. The telescope is equipped with the Yunnan Faint Object Spectrograph and Camera (YFOSC), which is a versatile instrument usable both for photometry and spectroscopy. An e2v back-illuminated $2048 \times 4608$ pixels CCD is mounted in YFOSC and covers a field of view of $10\arcmin \times 10\arcmin$ (with a pixel size of 0\farcs283\,pixel$^{-1}$) in the imaging mode. While the Lijiang telescope was primarily used for spectroscopy, images in the $V$ filter were also obtained as part of the program. Observations span from 22 October 2018 to 21 June 2019. The typical exposure time is 120 -- 150\,s (three 40 -- 50\,s consecutive exposures in each of the nights).

\section{Data Reduction} \label{sec:data}

\subsection{Swift}

The X-ray lightcurve is produced using the {\it Swift}/XRT data products generator\footnote{\url{https://www.swift.ac.uk/user_objects/}} \citep{evans07,evans09}.  We used this to extract the background-subtracted count rate of Mrk 142 in the 0.3 -- 10 keV energy range for each {\it Swift} snapshot during the campaign.

The \textit{Swift}/UVOT \citep{poole08} data analysis largely follows the same procedure detailed in \citet{edelson15,edelson17,edelson19} and is only described briefly here, focusing on details that differ. The data were processed using HEASOFT v6.24.  In the present study, field stars from the GAIA DR2 catalog \citep{GAIA_DR2} are used to refine the astrometry of each exposure before making photometric measurements. For each epoch and filter, fluxes are measured using the tool UVOTSOURCE.  Source extractions are measured using a circular region with a radius of 5\arcsec, while the background is measured in an annulus from 40--90\arcsec\ from which small circular regions centered on background stars are excluded (consequently, the background region resembles a ring of Swiss cheese, with holes of radius 12\arcsec\ centered on sources from the GAIA DR2 catalog that lie within 102\arcsec\ of Mrk~142).

The standard pipeline processing includes a correction for the gradual decline in UVOT sensitivity; the correction applied to the Mrk~142 data is an updated version that has been approved by the instrument team but has not yet (as of Summer 2019) been released in the CALDB (Alice Breeveld, private communication).  The data are then screened to identify which measurements are likely to be affected by detector regions with reduced sensitivity, applying the updated masks presented in Hern{\'a}ndez Santisteban et al. (submitted).  Any observation where Mrk~142 is identified as falling within the detector mask is then removed from the light curve.  Note that the masks used here do not include a correction for the UVOT shift-and-add processing, as the impact of this was not recognized until after the Mrk~142 data were analyzed. The result is that the masks of Hern{\'a}ndez Santisteban et al. are effectively smoothed by a blur on the scale of 7--10\arcsec, which causes a few false positive and false negative errors when screening the Mrk~142 measurements.

\subsection{Ground-based optical photometry}

The optical lightcurves were obtained using relative photometry.  For each exposure, the count rate within a circular aperture was obtained for Mrk~142 and a number of comparison stars.  Aperture radii for each telescope were:  11 pixels (3\farcs3) for OGG; 13 pixels (5\farcs1) for ELP; 7 pixels (2\farcs1) for the Liverpool Telescope and 5 pixels (4\farcs5) for the Zowada Observatory.  Background rates were extracted from an annulus with the following inner and outer radii: 40 to 60 pixels for OGG and ELP; 15 pixels to 20 pixels for the Liverpool Telescope and 20  to 30 pixels for Zowada Observatory.  These were optimized to maximize the signal to noise ratio (S/N) for each telescope.  Any differences in host galaxy contribution with changes in aperture size are corrected for by the inter-calibration of the lightcurves (described later).

For a given telescope, the fluxes for all exposures taken within 3 hours of each other were averaged, to improve S/N.  At each epoch, relative photometry is performed by dividing the observed rates by the sum of count rates from the chosen comparison stars.  We assess the reliability of the relative photometry through looking at the fractional standard deviation of the comparison stars used.  We experimented with the choice of how many comparison stars, and which ones to use as well as the aperture size.  We find that the choice of comparison stars does not affect the overall shape of the AGN light curve, but does have an important impact on the S/N.  The comparison stars used to produce the final lightcurves were chosen to give the lowest fractional standard deviation.  The comparison stars are between a factor of 2 -- 4 brighter than the AGN in the {\it g, r, i} and $z$ bands.  We add, in quadrature, the largest fractional standard deviation from the selected comparison stars to the statistical uncertainty in the relative AGN flux (though note that the fractional standard deviation is comparable for each of the comparison stars).   For the {\it g, r, i}, and $z$ filters we use the same 4 comparison stars for all detectors.  However, since the throughput in the $u$ band is much lower, and most stars are typically redder, we found more reliable photometry from choosing a different set of comparison stars for this band.  Despite this, the field of view of the OGG detector is significantly smaller than the other telescopes, forcing us to use a different set of comparison stars for the $u$ band.  Again, we find this does not change the shape of the lightcurve, and only effects the S/N.  The systematic and statistical uncertainties are of the same order, and on average we get better than 1\% photometry in the {\it g, r}, and $i$ filters for all telescopes/detectors.  In the $u$ filter the mean uncertainty is 2.4\%.  In the $z$ filter it varies by telescope/detector (see more below).  

The Lijiang 2-m data were analyzed separately, but also using relative photometry.  That analysis made use of 3 comparison stars, with a circular aperture of radius 9\farcs9.  The background rate was extracted from an annulus from 11\farcs3 to 14\farcs1.

\begin{figure}
\centering
\includegraphics[width=0.99\columnwidth]{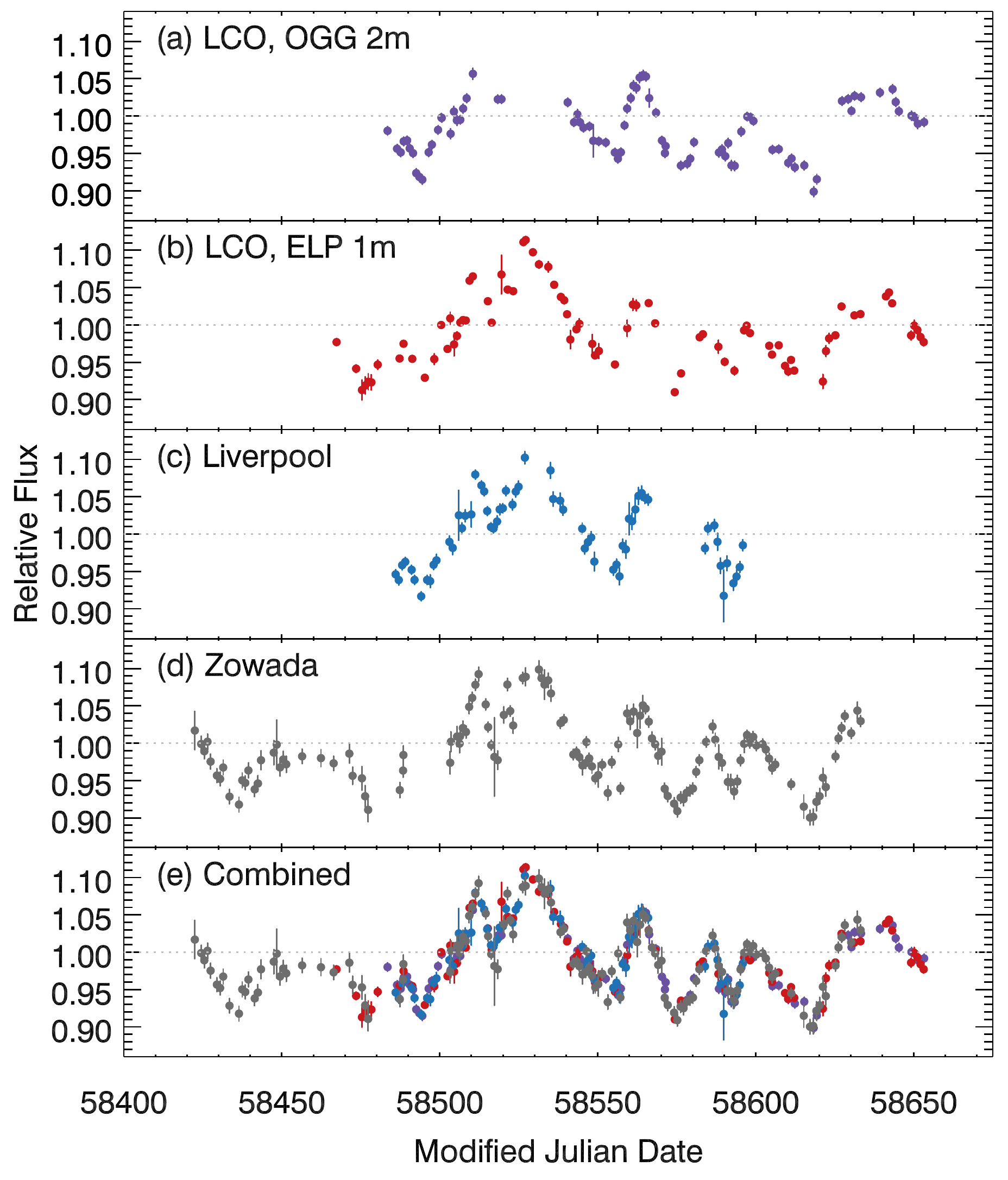}
\caption{$g$-band light curves of Mrk 142 from each of the ground-based telescopes.  Panels (a) and (b) show the LCO light curves from the OGG (purple) and ELP (red) telescopes, panel (c) shows the Liverpool Telescope (blue) light curve, while panel (d) shows the Zowada Observatory (gray) light curve.  Panel (e) shows the combined light curve from all telescopes,}
\label{fig:scalelc}
\end{figure}

Combining the lightcurves from each telescope and detector requires adding small shifts and scaling of the individual light curves to account for differences in bandpass and sensitivity for each combination of telescope and detector for a given filter.  In order to perform this intercalibration of the light curves we use the Bayesian method described by \citet{li14}\footnote{The code is publicly available here: \url{https://github.com/LiyrAstroph/CALI}}.  This method fits a damped-random walk model to all datasets simultaneously, allowing for a shift and scaling of each dataset in order to optimize the intercalibration.  It also takes into account the uncertainties on the best-fitting shift and scale parameters, increasing the uncertainties on the data points accordingly.  We show an example of the separate and combined light curves for the $g$-band only in Fig.~\ref{fig:scalelc}.   

\begin{figure*}
\centering
\includegraphics[width=0.85\textwidth]{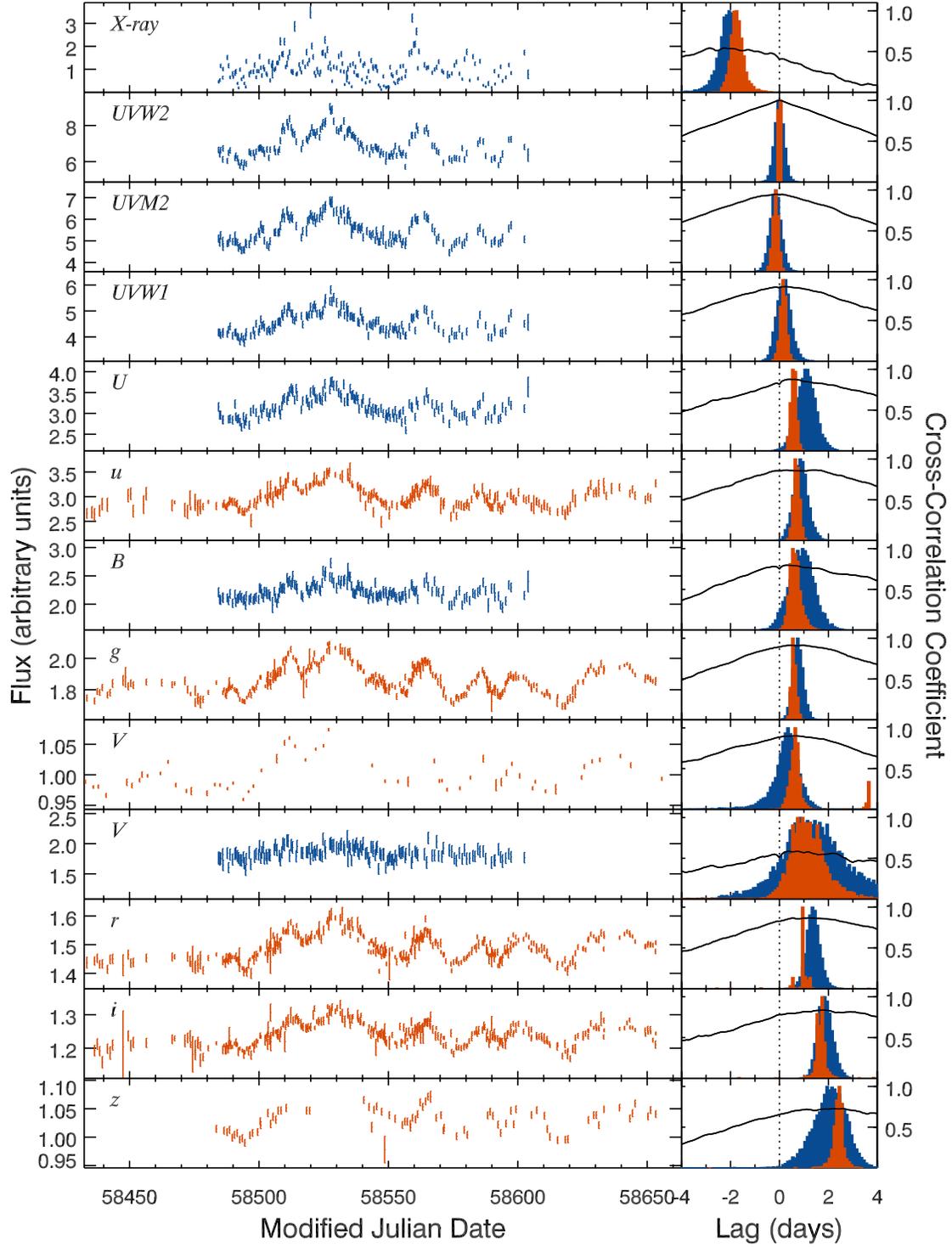}
\caption{{\it Left:} {\it Swift} (blue) and ground-based (orange) lightcurves of Mrk~142 during the 2019 monitoring campaign from shortest wavelength (top) to longest wavelength (bottom). {\it Swift} X-ray data are given as count rates, the {\it Swift}/UVOT and ground-based fluxes are given with units $10^{-15}$~erg~cm$^{-2}$~s$^{-1}$~\AA$^{-1}$, The exception is the Lijiang V light curve, which is given in arbitrary flux units normalized to a mean of 1.  {\it Right:} The solid black line shows the cross-correlation function between each waveband and the {\it UVW2} lightcurve.  Histograms show the probability distributions from ICCF (blue) and {\sc Javelin} (orange) lag measurements, with lags calculated with respect to the {\it UVW2} band.}
\label{fig:lc}
\end{figure*}

The mean uncertainty in the $z$ band is typically 1.5\%, however, this longest-wavelength band is also where the variability amplitude is lowest and on par with the flux uncertainty.  When performing the time-lag analysis (see Section~\ref{sec:timeseries}) we find that including all telescopes/bands combined gives significant scatter and leads to a poorly constrained lag measurement.  We explore the lags from the lightcurves from each telescope separately, finding that they are all consistent within 1$\sigma$, however, all are poorly constrained aside from the OGG $z$-band lightcurve.  We therefore opt to use only the OGG $z$-band lightcurve in all subsequent analysis since it is the highest quality (0.6\% mean photometric uncertainty) and thus provides the best-constrained lag measurement alone.  The light curves for all wavebands can be seen in Fig.~\ref{fig:lc}.  All lightcurves are given in Tab.~\ref{tab:lc}.

\begin{deluxetable*}{ccccc}
\label{tab:summary}
\tablecaption{Summary of observations used}
\tablewidth{0pt}
\tablehead{
 \colhead{Filter} & \colhead{Telescope} & \colhead{Date range} & \colhead{No. of epochs} & \colhead{Mean sampling rate}  \\
 &   & (MJD) & & (obs. per day)
 }
\decimalcolnumbers
\startdata
 X-ray  & {\it Swift} & 58484.3 -- 58603.9 & 185 & 1.55   \\ 
{\it UVW2} & {\it Swift} & 58484.3 -- 58603.9 & 149 & 1.25  \\
{\it UVM2} & {\it Swift} & 58484.4 -- 58602.5 & 146 & 1.24 \\
{\it UVW1} & {\it Swift} & 58484.3 -- 58603.9 & 151 & 1.26  \\
 $U$ & {\it Swift} & 58484.3 -- 58603.9 & 154 & 1.29  \\
 $B$ & {\it Swift} & 58484.3 -- 58603.9 & 168 & 1.40  \\
 $V$ & {\it Swift} & 58484.4 -- 58602.5 & 159 & 1.35  \\
 $V$ & Lijiang & 58413.9 -- 58655.6  & 62 & 0.26   \\
 $u$ & Liverpool & 58486.0 -- 58595.9  & 60  & 0.55  \\
    & LCO, OGG & 58483.5 -- 58653.3 & 78 & 0.46 \\
    & LCO, ELP & 58467.3 --  58555.4 & 42 & 0.48 \\
    & Zowada & 58422.4 -- 58636.2 &  107 & 0.50 \\
    & All & 58422.4 -- 58653.3 & 287 & 1.24   \\
 $g$ & Liverpool & 58486.0 -- 58595.9 & 63 & 0.57 \\
    & LCO, OGG &  58483.5 -- 58653.3 & 84 & 0.49 \\
    & LCO, ELP & 58467.3 -- 58653.1 & 80 & 0.43 \\
    & Zowada & 58422.5 --  58633.2 & 134 & 0.64 \\
    & All & 58422.5 --  58653.3 & 361 & 1.56  \\
 $r$ & Liverpool & 58486.0 -- 58595.9 & 62 & 0.56 \\
    & LCO, OGG & 58483.5 -- 58653.3 & 84 & 0.49 \\
    & LCO, ELP & 58467.3 -- 58653.1 & 78 & 0.42 \\	
    & Zowada & 58424.5 -- 58633.2  & 134 & 0.64  \\
    & All & 58424.5 --  58653.3 & 358 & 1.56  \\
 $i$ & Liverpool & 58486.0 -- 58595.9  & 63 & 0.57 \\
    & LCO, OGG & 58483.5 -- 58653.3 & 84 & 0.49  \\
    & LCO, ELP & 58467.3 -- 58653.1 & 77 & 0.41  \\
    & Zowada & 58424.5 -- 58633.2 & 131 & 0.63 \\
    & All & 58424.5 -- 58653.3 & 355 & 1.55  \\
 $z$  & LCO, OGG & 58483.5 -- 58653.3 & 81 & 0.48\\    
\enddata
\end{deluxetable*}

\begin{deluxetable}{ccccc}
\label{tab:lc}
\tablecaption{Mrk 142 lightcurves}
\tablewidth{0pt}
\tablehead{
 \colhead{Modified Julian Date} & \colhead{Filter} & \colhead{Rate} & \colhead{Uncertainty} & \colhead{Telescope}  }
\decimalcolnumbers
\startdata
58484.348   &   X-ray &  0.096 &      0.011 & Swift \\
58484.814   &  X-ray  &   0.111  &      0.012 & Swift\\
58485.139   &    X-ray & 0.241  &      0.017 & Swift\\
58485.934   &  X-ray  &  0.134   &      0.014 & Swift\\
58486.065   & X-ray  &   0.212   &      0.015 & Swift\\
\enddata
\tablecomments{This table is published online in its entirety in the machine-readable format.  A portion is shown here for guidance regarding its form and content. The X-ray rates are given as count rates, while the {\it Swift}/UVOT and ground-based fluxes have units of $10^{-15}$~erg~cm$^{-2}$~s$^{-1}$ \AA$^{-1}$. The exception is the Lijiang V light curve, which is given in arbitrary flux units normalized to a mean of~1.}
\end{deluxetable}

\section{Time Series Analysis} \label{sec:timeseries}

The lightcurves all show significant variability and are correlated -- the same prominent structures (peaks and troughs) can generally be seen at all wavelengths.  We therefore proceed to measure the time lags between the different wavebands.  We measure all time lags with respect to the {\it Swift}/{\it UVW2} light curve.  As the shortest-wavelength UV/optical band, with the highest variability amplitude (see variability amplitudes, $F_{\rm var}$, in Tab.~\ref{tab:lags}), {\it UVW2} is the natural choice for the reference band.  There are two methods typically used to measure lags between UV/optical AGN lightcurves: the interpolated cross-correlation function (ICCF) combined with flux randomization and random subset sampling  \citep[FR/RSS; as implemented by][]{petersonetal04} and the {\sc Javelin} analysis package \citep{zu11,zu13}.  Several recent works noted that uncertainties determined by ICCF were approximately two times larger than those determined by {\sc Javelin} \citep[e.g.][]{edelson19}. This motivated \citet{yu19}, who performed a detailed comparison of the two methods through extensive simulations.  Their conclusion was that {\sc Javelin} generally produces a more realistic estimate of the uncertainties than the ICCF method.  Here, we present lags and uncertainties determined via both methods, which we briefly describe in more detail.

For the ICCF method, we create many realizations of each of the lightcurves following the flux randomization and random subset sampling approach.  The data points in the lightcurve are randomly selected with replacement, meaning that some points are selected multiple times while others are not selected at all.  Error bars for data points are scaled appropriately for the number of times they are selected. Gaussian noise is then added to the data, with a mean equal to the observed flux, and standard deviation equal to the error bar.  The cross-correlation function of the realization is then calculated by linearly interpolating one lightcurve, then the other,  and averaging the two CCFs.  The peak and centroid of the CCF is then determined.  The centroid is calculated using CCF values higher than 80\% of the peak value.  This process is repeated $N=10,000$ times, leading to CCF centroid and peak distributions from which the median and  uncertainties are determined (using the 16\% and 84\% quantiles). 

{\sc Javelin} models the variability of  the lightcurves assuming a damped random walk prior constrained by the observed fluxes, and that the responding lightcurve is a delayed, blurred version of the reference lightcurve.  The transfer function connecting the two lightcurves is assumed to be a top-hat function.  {\sc Javelin} fits the lightcurves using a Markov Chain Monte Carlo algorithm, recovering the probability density distribution for the lightcurve and transfer function parameters.  We limit the lags to be within $-10$ to $+10$ days but otherwise run {\sc Javelin} with the default parameters.

The lags measured from both methods are given in Tab.~\ref{tab:lags} and are quoted in the observed frame.  While we quote both the peak and centroid lags from the ICCF method, we only use the centroid lags in the following analysis.  We also give the top-hat width from the {\sc Javelin} fits.  The right-hand panels of Fig.~\ref{fig:lc} show the CCFs as well as the lag distributions determined from both methods.  Note that we determine the uncertainty in the reference band by calculating the lag of the {\it UVW2} light curve with respect to itself.  Tab.~\ref{tab:lags} also gives the fractional variability amplitude, $F_{\rm var}$ \citep{vaughan03}, and the maximum correlation coefficient, $R_{\rm max}$, between the lightcurve of interest and the reference {\it UVW2} lightcurve.  The lags generally increase with wavelength rising from $<$1 days in the UV bands to $\sim$1.7--2.4 days in the $i/z$ bands (discussed more below).  We note that the UVM2 lag is slightly negative (though consistent with zero within 1$\sigma$).  The expected lag there is very small given the closeness in wavelength of the two filters, but we also note that the UVW2 filter has a larger red wing \citep{poole08} than the UVM2 filter, and thus may be more contaminated by longer wavelength light.  The UVM2 lag (with respect to UVW2) is always observed to be consistent with zero within 1$\sigma$ \citep[e.g., see Tab.~3 in][]{edelson19}.

The UV/optical lightcurves (aside from the {\it Swift/V} band) are well-correlated with the {\it UVW2} band (with $R_{\rm max} > 0.73$), and the X-ray lightcurve is the least well correlated of all the bands with $R_{\rm max} = 0.54$.  The {\it Swift/V} correlation is poor because the lightcurve is noisy -- the uncertainties on the data points are approximately the same size as the variability amplitude.  On the other hand, the X-ray lightcurve is poorly correlated with the {\it UVW2} because the well-measured rapid and large amplitude variations in the X-rays are absent from the {\it UVW2} lightcurve.

We explore this further by smoothing the X-ray lightcurve using a boxcar average and then recalculating the CCF with respect to the UVW2 lightcurve.  We vary the width of the boxcar from 1 to 10 days and re-evaluate $R_{\rm max}$ and the lag.  We find that smoothing significantly increases $R_{\rm max}$, with the strongest correlation of $R_{\rm max} = 0.74$ occurring with a boxcar width of 5 days. However, the smoothing does not significantly alter the UVW2 to X-ray lag, with the lag remaining consistent within 1$\sigma$.

Finally, we test splitting the {\it Swift} X-ray lightcurve up into soft (0.3 -- 1.5 keV) and hard (1.5 -- 10 keV) energies, and find that the {\it UVW2} lags of the two bands are consistent within 1$\sigma$, and therefore we do not explore these separate energy bands any further.

\begin{deluxetable*}{lcCCCCcc}
\label{tab:lags}
\tablecaption{Time lags calculated with respect to {\it UVW2} (observed frame), along with the variability amplitude and maximum correlation coefficient.}
\tablewidth{0pt}
\tablehead{
 \colhead{Filter} & Effective $\lambda$ & \colhead{$\tau_{\rm cent}~{\rm (days)}$} & \colhead{$\tau_{\rm peak}~{\rm (days)}$} & \colhead{$\tau~{\rm (days)}$} & \colhead{Top-hat width (days)} & \colhead{$F_{\rm var}$} & \colhead{$R_{\rm max}$} \\
  &  & {\rm ICCF} & {\rm ICCF} & \textsc{Javelin}& \textsc{Javelin}}
\decimalcolnumbers
\startdata
 X-ray & 0.3 -- 10 keV & -2.09_{-0.40}^{+0.34} & -2.2_{-0.6}^{+0.9} &  -1.74_{-0.23}^{+0.27} & 4.93_{-0.53}^{+0.50} & $0.533\pm0.007$ & 0.54 \\
 {\it UVW2} & 1928\,\AA &  0.00\pm0.20 & 0.0\pm0.1 & 0.00\pm0.01 & 0.07_{-0.05}^{+0.08} & $0.101\pm0.002$ & 1.00 \\
 {\it UVM2} & 2236\,\AA &  -0.15\pm0.24 & -0.2_{-0.1}^{+0.4} & -0.15\pm0.12 & 1.57_{-0.54}^{+0.51} & $0.097\pm0.003$ & 0.95\\
 {\it UVW1} & 2600\,\AA  &  0.18\pm0.29 & 0.3_{-0.5}^{+0.1} & 0.16\pm0.13 & 1.29_{-0.84}^{+0.94}  & $0.084\pm0.003$ & 0.92\\
 Swift, $U$ & 3467\,\AA &  1.09\pm0.39 & 0.7_{-0.3}^{+0.1} & 0.59\pm0.11 & 0.24_{-0.17}^{+0.34} & $0.073\pm0.003$ & 0.88\\
 $u$ & 3540\,\AA &  0.87\pm0.30 & 0.8_{-0.6}^{+0.7} &0.69_{-0.12}^{+0.13} & 3.58_{-0.45}^{+0.48} & $0.062 \pm 0.002$ & 0.87 \\
 Swift, $B$ & 4392\,\AA &  0.92_{-0.53}^{+0.50} & 0.3_{-0.5}^{+0.9} & 0.62_{-0.19}^{+0.24} & 1.32_{-0.88}^{+1.44} & $0.051\pm0.003$ & 0.80 \\
 $g$ & 4770\,\AA & 0.75_{-0.20}^{+0.23}  & 0.6_{-0.2}^{+0.4} & 0.58_{-0.09}^{+0.11} & 1.70_{-0.28}^{+0.48} &  $0.044 \pm 0.001$ & 0.91 \\
 $V$ & 5383\,\AA  & 0.32_{-0.54}^{+0.44} & 0.6_{-0.4}^{+0.3} & 0.66_{-0.20}^{+0.24} & 0.32_{-0.20}^{+0.44} & $0.025\pm0.001$ & 0.90 \\
 Swift, $V$ & 5468\,\AA & 1.40_{-1.20}^{+1.50} & 0.9_{-0.7}^{+1.5} & 1.14_{-0.61}^{+0.74} & 2.99_{-2.22}^{+4.02} &  $0.021\pm0.008$ & 0.59\\
 $r$ & 6215\,\AA & 1.38\pm0.27 & 1.2_{-0.6}^{+0.5} & 0.97_{-0.06}^{+0.18} & 0.009_{-0.006}^{+0.03} & $0.028\pm0.001$ & 0.87\\
 $i$ & 7545\,\AA &  1.85_{-0.29}^{+0.33} & 1.7_{-0.4}^{+0.6} & 1.69_{-0.17}^{+0.13} & 0.05_{-0.04}^{+5.97} &  $0.024\pm0.001$  & 0.84\\
 $z$ & 8700\,\AA & 2.08_{-0.73}^{+0.64} & 2.0_{-1.0}^{+0.8} & 2.42_{-0.19}^{+0.15}  & 0.31_{-0.26}^{+6.20}  & $0.019\pm0.001$ & 0.73 \\
\enddata
\end{deluxetable*}

\subsection{Lag-wavelength relation}

\begin{figure*}
\centering
\includegraphics[width=\textwidth]{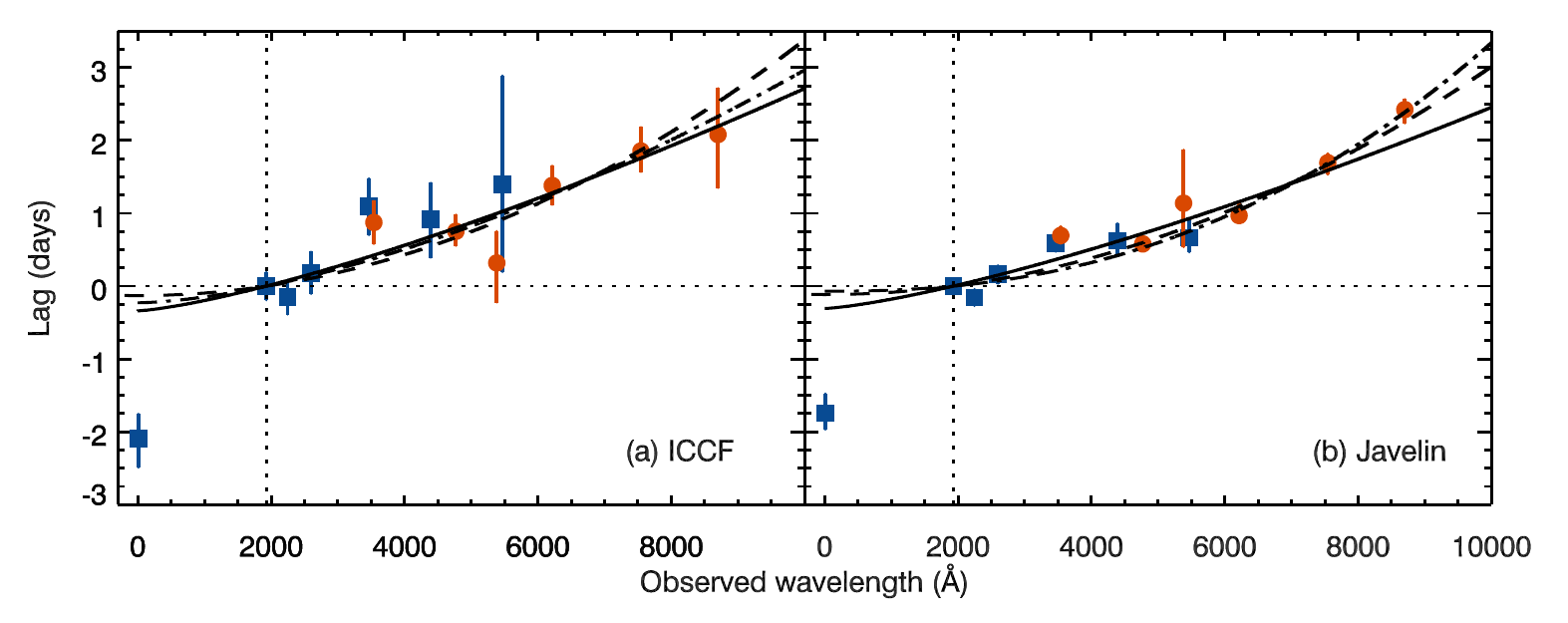}
\caption{Lag vs wavelength for Mrk 142 with lags determined by (a) ICCF and (b) {\sc Javelin}.  Lags shown are in the observed frame and are calculated with respect to the {\it UVW2} band which is indicated with a vertical dotted line (1928\AA).  {\it Swift} data points are given as blue squares, while lags from ground-based lightcurves are shown as orange circles.  The solid line shows the best-fitting $\tau \propto \lambda^{4/3}$ relation, the dashed line shows the best-fitting $\tau \propto \lambda^{2}$ relation, while the dash-dotted line shows the best-fitting $\lambda^\beta$ relation, where $\beta$ is a free parameter in the fit. } 
\label{fig:lagwave}
\end{figure*}

For a standard thin accretion disk the lags are expected to follow $\tau \propto \lambda^{4/3}$.  We therefore fit this relation to the observed lags in Mrk~142 using the following form:
\begin{equation}
\label{eq:lag}
\tau = \tau_0 \left[ \left(\lambda/\lambda_0\right)^\beta - 1.0 \right]
\end{equation}
where $\lambda_0 = 1928$\,\AA\ (the wavelength of the UVW2 band).  We  fit the relation both with $\beta = 4/3$ for a standard thin disk, $\beta = 2$ for a slim disk, and allowing $\beta$ to be a free parameter.  Initial fits show that the X-ray to {\it UVW2} lag is significantly offset from the best-fitting trend through the UV/optical, and therefore we remove the X-ray point from the fits.  Moreover, we find that the $u/U$ band lags also sit above the best-fitting relations (as has been seen in other objects), and therefore we also remove those points from the fits.  The best-fitting parameters are given in Tab.~\ref{tab:lagwave}, and are shown in Fig.~\ref{fig:lagwave}.  

The best-fitting slope is consistent with both $\beta = 4/3$ and $\beta=2$ for the ICCF lags, however, for the {\sc Javelin} lags a better fit is achieved with $\beta = 2$ than with $\beta = 4/3$.  The lag normalization parameter $\tau_0$ ranges from $\tau_0 = 0.07$ to $\tau_0 = 0.34$~days with the lower value from the {\sc Javelin} lags, which are systematically shorter than the ICCF lags (aside from the Lijiang {\it V} band, and the {\it z} band).  Noticing these lower {\sc Javelin} lags we also investigate the width of the top-hat function, finding that in some cases the width is large. For instance, for the $u$ band we find a width of $\sim$4 days, meaning that a significant portion of the response has negative lags.  We experiment with modifying the {\sc Javelin} code to force positive time lags.  For the $u$ band this increases the median lag from 0.59 days to 0.88 days, closer to the ICCF value.  We do not pursue this further here.

The difference in slope determined by the ICCF and {\sc Javelin} lags seems to be dominated by the $z$ band lag.  The best-fit to the {\sc Javelin} lags with the slope fixed at $\beta=4/3$ goes significantly below the $z$ band lag. While generally the {\sc Javelin} lags are smaller than the ICCF lags, the $z$ band lag is larger.  The combination of these leads to a larger slope when fitting the {\sc Javelin} lags.  Excluding the $z$-band lag the fit with $\beta=4/3$ (fixed) significantly improves, giving an acceptable fit with $\chi^2 = 8.5$ for 8 degrees of freedom. Given the strong dependence of the fits on the $z$ band lag -- the light curve with the lowest variability amplitude, and a lower number of data points -- we do not put too much weight on the implied larger slope from the {\sc Javelin} fits.

Previous studies have found that de-trending the light curves can reduce/remove the X-ray offset \citep[e.g.,][]{mchardy14,mchardy18}.  We therefore explored de-trending the X-ray and UVW2 lightcurves using a linear fit, a quadratic fit, and a boxcar average (of various widths).  We find that none of the de-trending methods significantly change the X-ray to UVW2 lag.

\begin{deluxetable}{lCcc}
\label{tab:lagwave}
\tablecaption{Fits to the lag-wavelength relation}
\tablewidth{0pt}
\tablehead{
\colhead{Lag method} &  \colhead{$\tau_0$ (days)} & \colhead{$\beta$} & \colhead{$\chi^2$ (dof)} }
\decimalcolnumbers
\startdata
ICCF  & 0.34\pm0.04 & 4/3 (fixed) & 3.46 (9)\\
ICCF  & 0.13\pm0.01 & 2 (fixed) & 3.67 (9)\\
ICCF  & 0.23\pm0.23 & $1.60\pm0.68$ & 3.30 (8) \\
{\sc Javelin} & 0.31\pm0.01 & 4/3 (fixed) & 18.3 (9) \\
{\sc Javelin}  & 0.12\pm0.01 & 2 (fixed) & 6.91 (9) \\
{\sc Javelin}  & 0.07\pm0.03 & $2.36\pm0.28$ & 5.30 (8)\\
\enddata
\end{deluxetable}

\section{Spectral Analysis}\label{sec:spec}

The structure of the disk can also be tested through an analysis of the spectrum of the variable component of the light curves.  To perform spectral modeling we first flux-calibrated the ground-based light curves using the magnitudes of the comparison stars from the AAVSO Photometric All-Sky Survey DR10 \citep{henden18}  for the {\it u, g, r} and $i$ bands, and the SDSS catalog for the $z$ band.  The flux-calibrated lightcurves (Tab.~\ref{tab:lc}) were then corrected for Galactic absorption assuming $E(B-V) = 0.0136$ and the extinction law of \citet{cardelli89}, and shifted to the rest-frame flux.

We perform a modified version of the flux-flux analysis to separate the constant (galaxy) and variable (AGN) components \citep[e.g.][]{cackett07,starkey17,mchardy18}.  We fit the light curves using the following linear model:

\begin{equation}
f_\lambda (\lambda, t) = A_\lambda(\lambda) + R_\lambda(\lambda)X(t)
\label{eq:flfl}
\end{equation}

\begin{figure}
\centering
\includegraphics[width=\columnwidth]{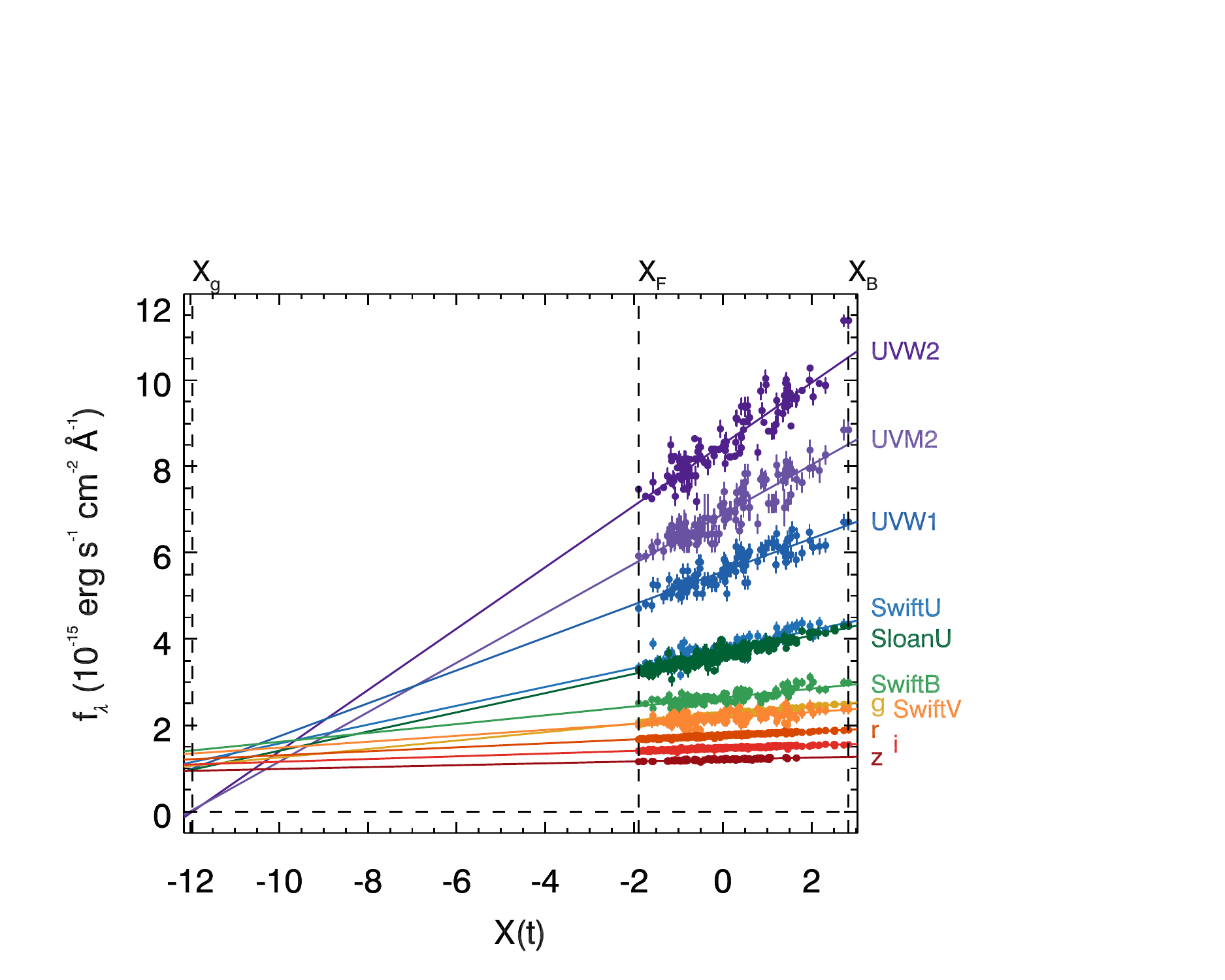}
\caption{Flux-flux analysis: $f_\lambda$ vs $X(t)$ for each of the light curves.  The best-fitting relations are shown as solid lines.  $X_F$ and $X_B$ indicate the faint and bright values of $X(t)$ during the campaign.  $X_g$ indicates the value of $X(t)$ where $f_\lambda = 0$ for the UVW2 band.  The flux-flux relations for other bands evaluated at $X_g$ give the minimum galaxy contribution.} 
\label{fig:fluxflux}
\end{figure}

Here $X(t)$ is a dimensionless light curve with a mean of 0 and standard deviation of 1.  $A_\lambda(\lambda)$ is a constant for each light curve, while $R_\lambda(\lambda)$ is the rms spectrum.  While this is a simplified model that does not take account of any time lags, the time lags only act to add scatter around the linear flux-flux relations.  We estimate the minimum host galaxy contribution in each band by extrapolating the best-fitting relations to where the first band crosses $f_\lambda = 0$.  In this case both the {\it UVW2} and {\it UVM2} bands cross $f_\lambda = 0$ at essentially the same value of $X$, which we denote $X_g$.  The host-galaxy components in the other bands are then the best-fitting relation evaluated at $X = X_g$.  Fig.~\ref{fig:fluxflux} shows the flux-flux relation ($X(t)$ vs $f_\lambda$) for each band.  Note that the linear relation in Eq.~\ref{eq:flfl} provides a good fit over the full range of observed fluxes in Fig.~\ref{fig:fluxflux}.  The absence of curvature here validates the assumption of a constant spectral shape for the variable light, and shows that any `bluer-when-brighter' effect in Mrk~142 is entirely due to a relatively blue spectrum of the variable light being diluted by a relatively red non-variable spectrum.

\begin{figure}
\centering
\includegraphics[width=\columnwidth]{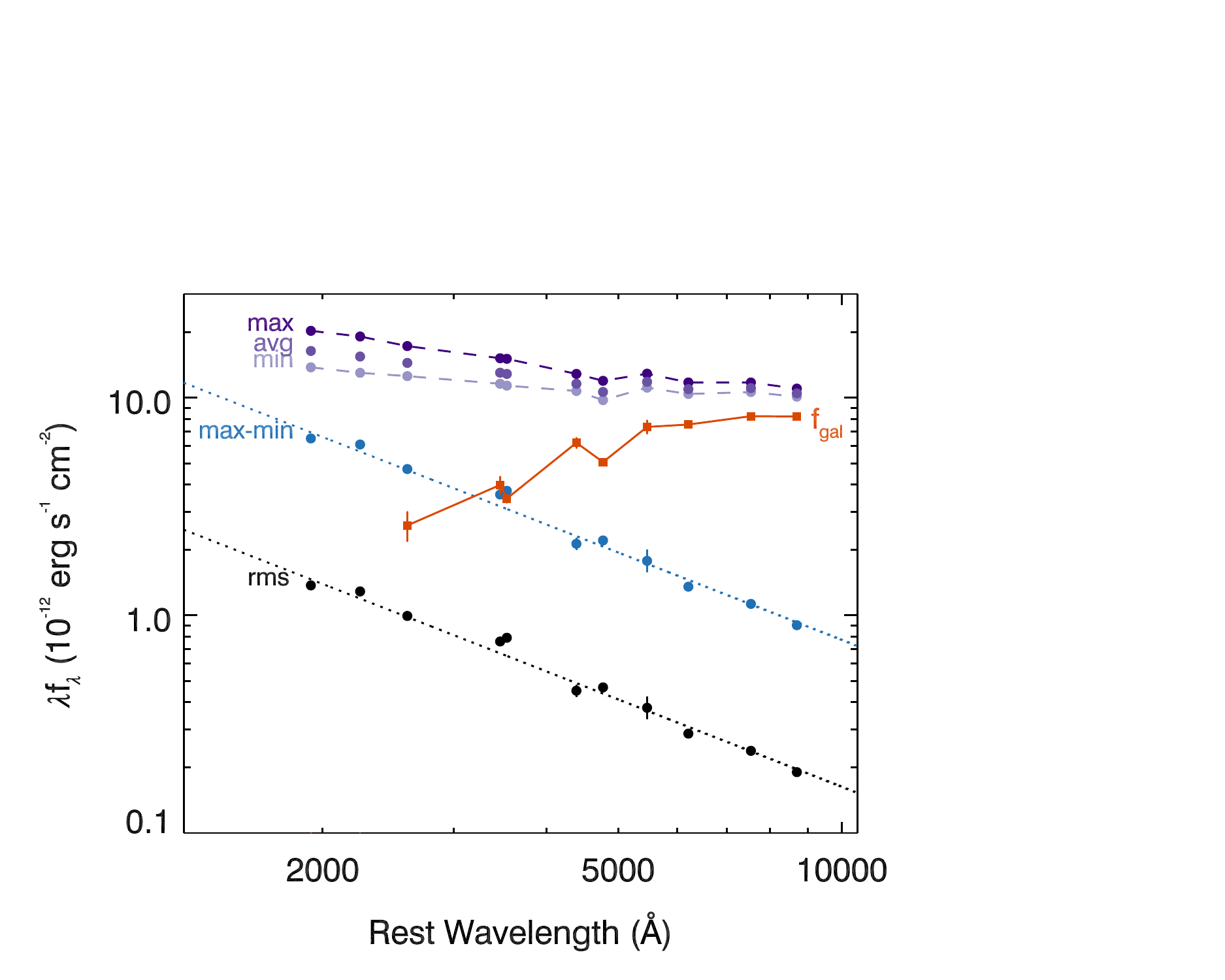}
\caption{The spectral energy distribution of Mrk 142 from the flux-flux analysis.  The maximum and minimum spectra are derived from the best-fitting relation evaluated at $X(t) = X_B$ and $X_F$, respectively, while the average spectrum comes from $X(t) = 0$.  The galaxy spectrum ($f_{\rm gal}$, orange squares) is derived from the best-fitting relation at $X(t) = X_g$.  The rms spectrum comes from the slope of the best-fitting relation.  The dotted lines indicate the best-fitting thin disk spectrum, $\lambda f_\lambda \propto \lambda^{-4/3}$, excluding the $U/u$ bands.  The variable spectrum of Mrk~142 is consistent with a standard thin accretion disk.} 
\label{fig:sed}
\end{figure}

In Fig.~\ref{fig:sed} we show the resulting spectral energy distributions.  The maximum and minimum spectra are determined from evaluating the best-fitting relations at the brightest and faintest values of $X = X_B$ and $X_F$ respectively.  The average spectrum is evaluated at $X = 0$.  The host galaxy components are shown as $f_{\rm gal}$ and generally increase with wavelength, as expected for an old stellar population.  The variable spectrum is plotted as both the maximum-minimum spectrum and the rms spectrum.  These variable spectra decrease with wavelength, and are well represented by the $\lambda f_\lambda \propto \lambda^{-4/3}$ relation expected for a standard thin disk (dotted lines in Fig.~\ref{fig:sed}).  If we allow the index to be a free parameter, we find $-1.36\pm0.01$, very close to the expected thin disk value of $-4/3$.  Note that a slim disk should have a spectrum of $\lambda f_\lambda =$~constant \citep{wang99b}, inconsistent with what we see here. Given the excess lags in the $U/u$ bands, we exclude those points from the spectral fits.  The $U/u$ fluxes lie 12 and 17\% above the best-fitting disk spectrum, putting additional constraints on contribution to the variable flux from the diffuse BLR.

\section{Discussion} \label{sec:discuss}

We have monitored the super-Eddington AGN Mrk~142 for 4 months with {\it Swift} with an average sampling rate of better than once per day.  Moreover, we obtained ground-based photometric monitoring in the Sloan {\it ugriz} filters over approximately 230 days, overlapping with the {\it Swift} monitoring.  By combining light curves from multiple telescopes around the globe (LCO, Liverpool, and Zowada) we obtain an average sampling rate of 1.6 observations per day in the $g$ band -- comparable to that obtained with {\it Swift}.

Mrk~142 was highly variable, with a variability amplitude as high as 53\% in the X-ray, dropping to 10\% in the {\it Swift}/{\it UVW2} band (1928\AA) and 1.9\% in $z$ ($\sim$9000\AA).  All the UV/optical bands are highly correlated with the {\it UVW2} light curve, with their maximum correlation coefficients all above 0.75 (aside from the noisy {\it Swift/V} band).  However, while the 0.3 -- 10 keV X-ray band shows a number of features that are apparent in the longer wavelength lightcurves, it shows much more variability on shorter timescales ($\sim$few days) that is not apparent at longer wavelengths, and has a significantly lower peak correlation coefficient of 0.54 (with respect to the {\it UVW2} band). Smoothing the X-ray lightcurve with a boxcar average of width 5 days removes the short timescale variability and leads to an increased peak correlation coefficient of 0.74 without affecting the lag measurement. While a significantly higher correlation, it remains weaker than correlations between the UVW2 and the highest quality UV/optical lightcurves.

The goal of the intensive photometric monitoring campaign was to perform the first continuum reverberation mapping of a super-Eddington AGN to test whether the accretion disk structure is notably different from that of previously-studied sub-Eddington AGN.  We therefore determine time lags between the {\it Swift}/UVW2 and other light curves.  We find that the lags increase with wavelength, approximately following $\tau \propto \lambda^{4/3}$, though they can also be fit with  $\tau \propto \lambda^{2}$ (but we caution that this is dependent on the $z$ band lag).  There are noticeable outliers, with the X-ray to UV time delay significantly longer than an extrapolation of the best-fit through the UV and optical.  Moreover, the $u/U$ lags are also significantly offset from the general trend with wavelength.  Rather surprisingly given the significantly higher mass accretion rate of Mrk~142, the main observational results -- that approximately $\tau \propto \lambda^{4/3}$; the X-ray offset and poor correlation with the UV/optical; and the enhanced $u/U$ lags -- are seen in all the other high-cadence {\it Swift} monitoring campaigns to date on sub-Eddington AGN \citep{edelson15,edelson17,edelson19,fausnaugh16,cackett18,mchardy18}.

The origin of the poor X-ray/UV correlation remains unclear.  In comparing NGC~5548, NGC~4151, NGC~4593 and Mrk~509, \citet{edelson19} noted that they all show a poorer correlation between the X-rays and the UV than the UV and the optical.  This is hard to reconcile with a picture where the X-rays directly irradiate the UV/optical part of the accretion disk driving the variability.  Even more puzzling are objects such as Mrk~817, where no correlation at all is seen between the X-rays and the UV \citep{morales19}, despite this being quite a typical object where broad emission line reverberation is observed.  \citet{gardnerdone17} explain the poor X-ray correlation in NGC~5548 through a vertically extended inner Comptonizing region that prevents the X-rays directly irradiating the disk.  Even then, light travel time from this inner Comptonizing region is too short to explain the lags, and thus they suggest the lags may instead be a dynamical timescale for the outer disk to respond to changing FUV illumination.  \citet{edelson17} invoke a similar inner torus to explain the long X-ray to UV lags in NGC~4151.  Here, in Mrk~142 this may also be a natural explanation -- given the high mass accretion rate, the inner disk is expected to be slim (not thin) within the photon trapping radius and vertically extended.  This might act in the way envisaged by \citet{gardnerdone17}.   However, while such an inner-disk structure is expected for slim-disk models for super-Eddington AGN it remains a puzzle as to why the sub-Eddington AGN exhibit the same phenomenon. 

The large X-ray to UV lag has important implications for the size of the BLR too.  If the X-rays are a good proxy for the driving light curve, then the 2 day lag between X-rays and  {\it UVW2} variations would imply that the BLR size in \citet{Du16} is underestimated by $\sim$40\%.  \citet{Du16} measure a H$\beta$ lag with respect to the 5100\AA\ continuum of approximately 8 days.  Since the {\it UVW2} to 5100\AA\ lag is approximately 1 day, and the X-ray to {\it UVW2} lag is about 2 days, the `true' H$\beta$ lag would be 11 days.  If, however, the {\it UVW2} band is closer to the driving continuum, then the BLR size is only underestimated by $\sim$10\% (8 vs. 9 days).  Therefore, it is important to develop a better understanding of what band is driving the optical variability, since it has implications for BLR size and hence black hole mass estimates.

The excess lag in the $U/u$ bands is thought to be due to continuum emission arising in the BLR.  This emission, which contributes to the observed continuum over a broad range in wavelengths from the UV to the near-IR, has a significant discontinuity at the Balmer jump (3646\AA) \citep{koristagoad01, lawther18,koristagoad19} and therefore leads to an increase in the lags particularly around that wavelength.  $U/u$ band excesses have been seen in NGC~5548 \citep{edelson15,fausnaugh16}, NGC~4151 \citep{edelson17}, NGC~4593 \citep{cackett18,mchardy18}, and Mrk 509 \citep{edelson19}.  The UV spectroscopic observations of NGC~4593 were particularly powerful in highlighting this, showing a broad excess in the lags around the Balmer jump \citep{cackett18}, rather than from just a single broadband photometric filter.  The $U/u$ band excess observed here in Mrk~142 likely has the same origin due to continuum emission from the BLR.   \citet{edelson19} compared the magnitude of the $U$ excesses in four objects, finding that on average the excess was a factor of 2.2 larger than expected from the best-fitting lag-wavelength relation, with values ranging from 1.6 to 2.9.  For Mrk~142 we find that the $U/u$ lags are on average a factor of 2.4 larger than the best-fitting lag-wavelength relation, consistent with the results of \citet{edelson19}. Properly assessing the impact of the BLR continuum on the lags requires careful spectral deconvolution and light curve simulations \citep[e.g.,][]{koristagoad19}, which should be possible for Mrk~142 once analysis of the optical spectra from this campaign is completed in the future.

Another consideration is how the normalization of the lag-wavelength relation compares to the expectations from assuming a standard Shakura-Sunyaev disk \citep{shakurasunyaev} with temperature profile $T \propto R^{-3/4}$.  To do this, we use Eq.~12 from \citet{fausnaugh16} for the normalization of the lag-wavelength relation, $\tau_0$, which we reproduce here:
\begin{equation}
\tau_0 = \frac{1}{c}\left( X \frac{k \lambda_0}{hc} \right)^{4/3} \left[ \left(\frac{GM}{8 \pi \sigma}\right)\left(\frac{L_{\rm Edd}}{\eta c^2}\right)(3 + \kappa)\dot{m}_{\rm E}\right]^{1/3} \; .
\label{eq:faus}
\end{equation}
In this equation $\eta$ is the accretion efficiency, $X$ is a factor for converting from $\lambda$ to $T$ for a given radius, $\kappa$ is the local ratio of external to internal heating, and  $\dot{m}_{\rm E} = L_{\rm bol}/L_{\rm Edd}$.  Here, we assume a flux-weighted value for $X = 2.49$ (though note response-weighted values will be larger), $\kappa = 1$ and a black hole mass of $M = 1.7\times10^6 $~M$_\odot$ \citep{li18}.  To compare with the observations we first consider several estimates for the bolometric luminosity, $L_{\rm bol}$, given that bolometric corrections can sometimes be highly uncertain.  First, we determine $L_{\rm bol}$ using $L_{\rm bol} = 9\lambda L_\lambda$ (5100\AA) \citep{kaspi00}.  Since we do not directly measure the 5100\AA\ flux, we use the {\it Swift} $V$-band flux as an estimate.  From the flux-flux analysis we determine an average host-galaxy subtracted rest-frame flux of $8.3\times10^{-16}$~erg~cm$^{-2}$~s$^{-1}$~\AA$^{-1}$.  This leads to $L_{\rm bol} = 1.85\times10^{44}$~erg~s$^{-1}$ for a luminosity distance of $D_L = 201.5$~Mpc, and $L_{\rm bol}/L_{\rm Edd} = 0.86$ (for a black hole mass of $M = 1.7\times10^6 $~M$_\odot$).  Alternatively, we can use the observed 2 -- 10 keV X-ray flux and the bolometric correction of \citet{marconi04}.  Using the average X-ray spectrum from {\it Swift} we measure a 2 -- 10 keV flux of $1.9\times10^{-12}$~erg~s$^{-1}$~cm$^{-2}$, which in turn leads to $L_{\rm bol} = 1.6\times10^{44}$~erg~s$^{-1}$ and $L_{\rm bol}/L_{\rm Edd} = 0.74$.  Finally, we can use the observed host galaxy-subtracted 5100\AA\ luminosity combined with the Shakura-Sunyaev disk model itself to estimate the dimensionless mass accretion rate $\dot{\mathscr{M}}$  following Eq.~2 in \citet{Du15}.  $\dot{\mathscr{M}}$  relates to the Eddington ratio via $L_{\rm bol}/L_{\rm Edd} = \eta \dot{ \mathscr{M}}$.  We get $\dot{\mathscr{M}} = 100$ during this campaign.  To convert to an Eddington ratio we must assume some accretion efficiency, but, this is expected to drop with increasing mass accretion rate for slim disk  models \citep{wang99,mineshige00,sadowski11}.  Using the formulation of \citet{mineshige00} we determine $\eta = 0.034$ and $L_{\rm bol}/L_{\rm Edd} = 3.4$ for $\dot{\mathscr{M}} = 100$.  Thus, the three estimates give a range of 0.74 to 3.4 for $L_{\rm bol}/L_{\rm Edd}$.

Using these estimates for $L_{\rm bol}/L_{\rm Edd}$ and $\eta$ we can now compare the observed and predicted values for $\tau_0$.  Under the assumptions above,  taking $L_{\rm bol}/L_{\rm Edd} = 3.4$, and $\eta = 0.034$ we predict $\tau_0 = 0.1$~days. In other words, the observed $\tau_0$ (assuming $\beta = 4/3$; see Tab.~\ref{tab:lagwave}) is a factor of 3.1 to 3.4 larger than predicted from the standard disk model and our largest estimate of $L_{\rm bol}/L_{\rm Edd}$.  This discrepancy between observed and predicted disk size is comparable to what is seen in other objects, \citep[e.g.,][and references therein]{edelson19}.  Either a significantly higher Eddington ratio, or lower accretion efficiency would be needed to reconcile the model lags, in other words, the magnitude of the lags is consistent with a highly super-Eddington accretion rate.  However, we recognize that since Eq.~\ref{eq:faus} is for a standard sub-Eddington disk it would no longer be applicable.

The discrepancy between the observed and predicted disk size is similar to the issue in sub-Eddington objects -- for reasonable accretion rates, the magnitude of the predicted lags is a factor of a few smaller than observed \citep[e.g.,][]{edelson15,edelson19,cackett18,mchardy14,mchardy18}. Solutions that have been proposed for sub-Eddington objects include inhomogeneous accretion disks \citep{dexter11}, a tilted inner disk \citep{starkey17}, that the lags are due to a dynamical timescale for the outer disk to respond to changing FUV illumination \citep{gardnerdone17}, that the X-ray source is located higher above the disk than usually assumed \citep{kammoun19}, or that the lags are due to disk turbulence \citep{cai20}.  Continuum emission from the BLR will also contribute, or even dominate, the observed lag \citep{koristagoad01,koristagoad19,lawther18,chelouche19}.  Those same solutions could work here also.  Alternatively, for higher-mass accretion rate objects there may be other solutions.  For instance, the model used for the lag-wavelength relation assumes a standard optically-thick geometrically-thin accretion disk, and so will be not applicable if the disk is instead a slim disk.  At high mass accretion rates the inner region of a slim disk is expected to be geometrically thick and will create an anisotropic radiation field that is not taken into account here.

Additonal tests of the disk structure can be performed through analysis of the variable spectrum.  Thus, we also performed a flux-flux analysis to decompose the observed spectrum into constant and variable components.  We found that the spectrum of the variable component is well-represented by $\lambda f_\lambda \propto \lambda^{-4/3}$, as expected for a standard thin disk.  The variable spectrum is not consistent with a slim disk.  The constant component increases with wavelength, as expected for stellar population.  The $U/u$ band fluxes are enhanced by a little over 10\% with respect to the best-fitting $\lambda^{-4/3}$ relation, which can be used to constrain any flux due to continuum emission from the BLR.

Since the variable spectrum is consistent with a thin disk (and rules out a slim disk in the UV/optical), and given that the UV/optical lags can be fit by $\tau \propto \lambda^{4/3}$, this has implications for the accretion disk structure at such high Eddington ratios.  In the slim disk model \citep[e.g.,][]{abramowicz88} the accretion disk increases in scale height within the photon trapping radius.  Observations of the BLR support this, with higher mass accretion rate objects falling significantly below the radius-luminosity relation, as would be expected if an inflated inner disk was shadowing it \citep{Du15,Du16,du18}.  Broad-line reverberation of Mrk~142 shows that it also falls below the radius-luminosity relation \citep{Du16}, suggesting it contains a slim disk. If the accretion disk in Mrk~142 is a slim disk, then our spectral analysis shows that since the optical/UV emitting part looks like a thin disk, then the inflated inner disk must be well within the region producing the UV emission we observe with the {\it Swift}/{\it UVW2} (1928\AA).  We can therefore put observational constraints on the size of the photon-trapping radius by assuming that the extrapolation of the UV/optical lags ($\tau_0$)  sets the maximum extent of the photon-trapping region.  For our largest $\tau_0$ estimate of 0.34 days this corresponds to the light travel time for a distance of $1\times10^{13}$\,m, or for a black hole mass of $M = 1.7\times10^6 $\,M$_\odot$, it corresponds to approximately $4\times10^3$\,$R_{\rm g}$  (where $R_{\rm g} = GM/c^2)$.  

From a theoretical perspective, according to the self-similar solution \citep{wang99}, the trapping radius is given by
\begin{equation}
\frac{R_{\rm tr}}{R_{\rm g}}=4.5\times 10^{2}\left(\frac{\dot{\mathscr{M}}}{250}\right),
\end{equation}
and the effective temperature is
\begin{equation}
T_{\rm eff}=9.4\times 10^{5}\left(\frac{M}{10^{6}M_{\odot}}\right)^{-1/4}
            \left(\frac{R}{R_{\rm g}}\right)^{-1/2},
\end{equation}
Using Wien's law to go from temperature to wavelength we have
\begin{equation}
\frac{R}{R_{\rm g}}=1.1\times 10^{3}\left(\frac{M}{10^{6}M_{\odot}}\right)^{-1/2}
                    \left(\frac{\lambda}{1000{\rm \AA}}\right)^{2},
\end{equation}
giving the trapping radius in terms of wavelength as: 
\begin{equation}
\lambda_{\rm tr}=0.65\times 10^{3}\left(\frac{M}{10^{6}M_{\odot}}\right)^{1/4}
                 \left(\frac{\dot{\mathscr{M}}}{250}\right)^{1/2}\,{\rm \AA}.
\end{equation}
These scaling relations show that $\tau\propto \lambda^{2}$ in the trapping region, which 
is steeper than the standard disk model.  While this is consistent with the lags we observe in Mrk~142, the spectrum of the variable component is far from the flat spectrum expected for a slim disk, and very well-fit by a standard thin disk spectrum.
Moreover, for Mrk~142 with $M=1.7\times 10^{6}$\,M$_{\odot}$ and $\dot{\mathscr{M}}\approx 100$ (from our estimate above),  the optical and UV photons are not trapped ($\lambda_{\rm tr} = 462$\AA), but the soft X-ray photons should be trapped.  This appears to be consistent with the X-ray offset and poor X-ray/UV correlation. However, as noted above, while an inner geometrically-thick region is expected for slim disks, it is not expected in standard geometrically-thin disks in sub-Eddington sources and thus it remains a puzzle as to why those sources also show an X-ray offset and poor X-ray/UV correlation.

In summary, the high cadence, multi-wavelength photometric monitoring of  Mrk~142 has provided a rare opportunity to place observational constraints on the accretion flow at super-Eddington rates.

\acknowledgements
EMC and JM gratefully acknowledge support for the {\it Swift} analysis from NASA through grant 80NSSC19K0150, and support for analysis of the ground-based data from the NSF through grant AST-1909199.  EMC and Wayne State University deeply thank the 419 Foundation for donating the Dan Zowada Memorial Observatory.  We are also grateful to Terry Friedrichsen, Philip Moores, Nick Paizis, and Dennis Recla for help with maintenance at the Zowada Observatory.  Without them we would not be able to operate as many nights as we do.   

JMG and RE gratefully acknowledge support from NASA under the ADAP award 80NSSC17K0126.  KH acknowledges support from STFC grant ST/R000824/1. Research by AJB is supported by NSF grant AST-1907290. LCH was supported by the National Science Foundation of China (11721303, 11991052) and the National Key R\&D Program of China (2016YFA0400702). JMW acknowledges financial support from the the National Science Foundation of China (11833008 and 11991054), from the National Key R\&D Program of China (2016YFA0400701), from the Key Research Program of Frontier Sciences of the Chinese Academy of Sciences (CAS; QYZDJ-SSW-SLH007), and from the CAS Key Research Program (KJZD-EW-M06). PD acknowledges financial support from the the National Science Foundation of China (11873048 and 11991051) and from the Strategic Priority Research Program of the CAS (XDB23010400). CH acknowledges financial support from the the National Science Foundation of China (11773029). YRL acknowledges financial support from the the National Science Foundation of China (11922304),from the Strategic Priority Research Program of the CAS (XDB23000000), and from the Youth Innovation Promotion Association CAS. BL acknowledges financial support from the National Science Foundation of China grant 11991053. YFY is  supported by National Natural Science Foundation of China (Grant No. 11725312, 11421303).  This work made use of data supplied by the UK Swift Science Data Centre at the University of Leicester.

\bibliographystyle{aasjournal}
\bibliography{agn}

\end{document}